\documentclass[aps,pra,twocolumn,groupedaddress]{revtex4-1}

\usepackage{graphicx}
\usepackage{latexsym}
\usepackage{bm,array}
\usepackage{amsfonts}
\usepackage{amssymb}
\usepackage{amsmath}
\usepackage{upgreek}

\newcommand{\Ai}{\operatorname{Ai}}

\begin{document}


\title{The Boltzmann distribution and the quantum-classical correspondence}


\author{Sam Alterman$^1$,
Jaeho Choi$^{1,2}$,
Rebecca Durst$^{1,3}$,
Sarah M.~Fleming$^1$,
William K.~Wootters$^1$}
\affiliation{$^1$Department of Physics, Williams College, Williamstown MA 01267 \\
           $^2$Department of Mathematics, University of Maine, Orono ME 04469 \\
           $^3$Division of Applied Mathematics, Brown University, Providence RI 02912}



\begin{abstract}
In this paper we explore the following question: can the probabilities constituting the quantum Boltzmann
distribution, $P^B_n \propto e^{-E_n/kT}$, be derived from a
requirement that the quantum configuration-space distribution for a system in thermal equilibrium be 
very similar to the corresponding classical distribution?  
It is certainly to be expected that the quantum distribution in configuration space will approach the classical distribution as the temperature approaches infinity, and a well-known equation derived from
the Boltzmann distribution shows
that this is generically the case.  Here we ask whether one can reason in the opposite direction, that is, from
quantum-classical agreement to the Boltzmann probabilities.
For two of the simple 
examples we consider---a particle
in a one-dimensional box and a simple harmonic oscillator---this 
approach leads to probability distributions that provably approach the Boltzmann probabilities
at high temperature, in the sense that the Kullback-Leibler divergence between the 
distributions approaches zero.
\end{abstract}

\pacs{}

\maketitle


\section{Introduction}
\label{intro}
Much has been written about the correspondence between quantum mechanics and classical mechanics.  Papers on the subject range from early identifications of similarities, 
such as the Ehrenfest theorem \cite{Ehrenfest} and
the phase-space representation of quantum mechanics \cite{Wigner,Weyl,Moyal}, to more recent efforts to understand the role of decoherence in the quantum-to-classical transition \cite{Zurek,Schlosshauer,Halliwell,Landsman} and to assess the status of the correspondence principle
in light of chaotic dynamics \cite{Landsman,Ikeda,Belot}.  In this paper we study a specific aspect of the quantum-classical correspondence, namely, the way in which the
quantum mechanical configuration-space distribution for a system in thermal
equilibrium approaches its classical counterpart as the temperature gets large.  
(The questions we raise here could also be raised for probability distributions over
other slices of phase space.  But in this paper we restrict our attention to the
configuration-space distribution.)

For each of the examples we consider here, which are all quite simple and in fact involve only a 
one-dimensional configuration space, we observe that the quantum distribution 
becomes quite similar to the classical distribution 
even at modest values of the temperature.  This similarity depends on a kind of coordination
between the shapes of the quantum mechanical energy eigenfunctions (in the position representation) and the Boltzmann weights with which the squares of these eigenfunctions are 
averaged in a thermal mixture: these two elements work together to produce
a high level of agreement with the classical distribution.   The degree of this agreement leads us to ask whether
the probabilities constituting the quantum Boltzmann distribution, $P^B_n \propto e^{-E_n/kT}$
(which together with the energy eigenstates define the canonical ensemble),
can be {\em deduced} from a requirement that the quantum position distribution be very similar
to the classical position distribution.  If such a deduction is indeed possible, it may indicate a tighter network of connections
among quantum mechanics, classical mechanics, and statistical mechanics than we normally recognize.  After all, we already have other, apparently independent ways of deriving the
Boltzmann distribution.    
 
We do not answer our question in general---it seems to be a difficult one.  Rather, we show how a 
derivation of the Boltzmann probabilities can be achieved in our examples, at least
for high temperatures, and we argue
that the matter is worth further investigation.  
Crucially, in trying to arrive at the probabilities $P^B_n$,
we do not put into the mathematics the energy eigenvalues $E_n$ but only the squares of the
energy eigenfunctions.  

To be sure, one {\em expects} the quantum distribution in configuration space to approach the 
corresponding classical distribution
as the temperature gets large.  One way to see this is through the following well-known approximate
expression for the quantum distribution (based on the Boltzmann formula), in which only those terms up to second order in $\hbar$
have been kept \cite{Oppenheim}
(the approximation is written here for the one-dimensional case):
\begin{equation}  \label{expansion}
\begin{split}
\rho_q(x) \propto &\left\{ 1 + \hbar^2 \left[ \left( - \frac{1}{12 mk^2T^2} \right) \frac{d^2 V(x)}{dx^2}\right.\right. \\
&\left. \left. +\, \frac{1}{24 mk^3T^3} \left(\frac{dV(x)}{dx}\right)^2 \right] \right\} e^{-V(x)/kT}.
\end{split}
\end{equation}
Here $V(x)$ is the potential energy and $m$ is the particle's mass.  
As the classical position distribution $\rho_c(x)$ is simply proportional to 
$e^{-V(x)/kT}$, Eq.~(\ref{expansion}) predicts that as the temperature approaches 
infinity, the ratio $\rho_q(x)/\rho_c(x)$ approaches the constant value 1.  We could use Eq.~(\ref{expansion}) or related equations
to try to account for the similarity we see between the quantum and classical cases \cite{fn1},
but our aim in this paper is not so much to account for this similarity as to {\em use} it as a starting point for generating the Boltzmann probabilities.  

In the following sections we consider three simple examples, again all in one spatial dimension: (i) a particle in a box with infinitely hard walls, (ii) a
harmonic oscillator, and (iii) a particle in a linear potential bounded by a hard
wall.  For each example, we begin by illustrating the similarity between the quantum and classical 
position distributions.  Then we ask this question: given the classical position distribution
and the squares of the quantum energy eigenfunctions, how should the latter be weighted
in order to produce a position distribution ``as close as possible'' to the classical one?  Or, slightly
more precisely, if we let $\rho(x)$ be the position distribution resulting from the weighting of the
squared eigenfunctions, 
we ask how the weights should be chosen so as to make the
{\em ratio} $r(x) = \rho(x)/\rho_c(x)$ ``as flat as possible,'' where $\rho_c(x)$ is the classical thermal
distribution at some temperature $T$.  
We particularly want to know whether the optimal weighting is similar to the Boltzmann
weighting, and whether it approaches the Boltzmann weighting
as the temperature gets large \cite{newfootnote}.

Ideally we would adopt a single, simple interpretation of ``as flat as possible'' that applies
to all our examples, but we have
not found such an interpretation that is mathematically tractable.  Instead, in each case 
we interpret ``as flat as possible'' in a way that makes the mathematics manageable for that
example.  These {\em ad hoc} interpretations are, however, sufficiently similar to each other that
they can guide us, in Section V, to a general definition of ``as flat as possible'' that we can reasonably hope will generate the Boltzmann probabilities, in the high-temperature limit, for a large class
of systems.  

The final section then discusses the potential significance of our observations.

\section{Particle in a box}
\label{sec:2}
Consider a one-dimensional particle of mass $m$ confined to a box with hard walls at the positions $x=0$ and $x=L$.  
The quantum energy eigenvalues for this system are $E_n = \pi^2 \hbar^2n^2/(2 m L^2)$, where
$n = 1, 2, \ldots$.  
In terms of a rescaled position $\xi = x/L$ and a
rescaled temperature $\tau = kT/E_1$,  the energy eigenfunctions are
\begin{equation}
\psi_n(\xi) = \sqrt{2}\sin(\pi n \xi),  \hspace{2mm} n = 1, 2, \ldots ,
\end{equation}
and the Boltzmann probabilities with which these eigenfunctions are weighted in thermal equilibrium are
\begin{equation} \label{pBs}
P^B_n = \frac{1}{Z}e^{-n^2/\tau},
\end{equation}
where $Z = \sum_{n=1}^{\infty} e^{-n^2/\tau}$.  Thus, in thermal equilibrium, the quantum distribution of the particle's position
is 
\begin{equation}
\begin{split}
\rho_q(\xi) &= \sum_{n=1}^{\infty} P^B_n |\psi_n(\xi)|^2 = 1 - \sum_{n=1}^{\infty} P^B_n \cos(2 \pi n \xi) \\
&=\frac{\theta_3(0,e^{-1/\tau})-\theta_3(\pi \xi, e^{-1/\tau})}{\theta_3(0,e^{-1/\tau})-1},
\end{split}
\end{equation}
where $\theta_3$ is the Jacobi theta function $\theta_3(z,q) = \sum_{n=-\infty}^\infty 
q^{n^2}e^{2niz}$.
Meanwhile the analogous {\em classical} distribution of the particle's position, at any non-zero 
temperature, is simply the uniform distribution (because inside the box, the potential energy
does not depend on position):
\begin{equation}
\rho_c(\xi) = 1.
\end{equation}

\begin{figure}[h]
\begin{center}
\includegraphics[scale=0.9]{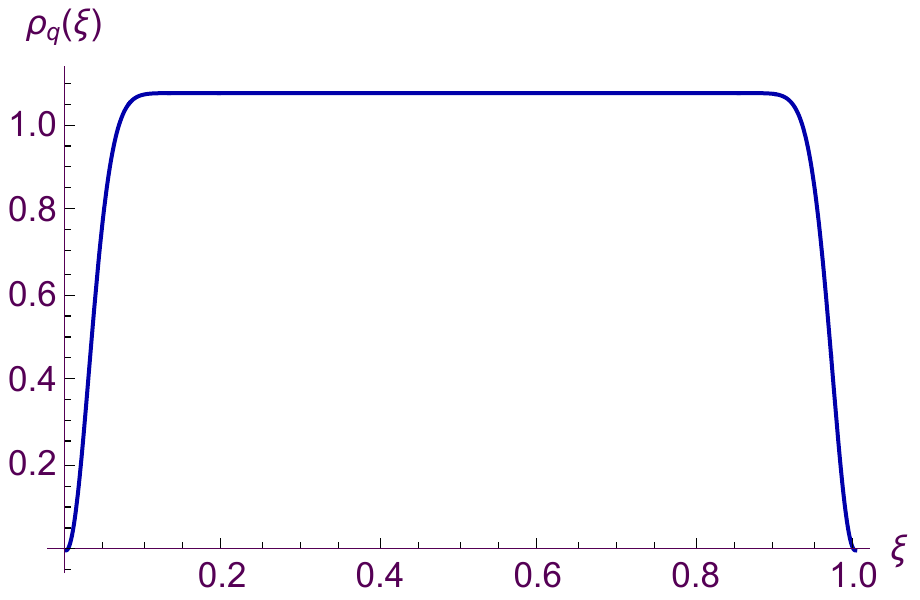}
\end{center}
%
%
\caption{The thermal position distribution of a quantum particle in a box.  Here $kT = 60E_1.$}
\label{boxpos}       
\end{figure}

Fig.~\ref{boxpos} plots the quantum position distribution $\rho_q(\xi)$ for the temperature
$\tau = 60$.  
This is not a particularly high temperature: the first fourteen energy eigenstates
carry 99\% of the probability.  And yet the distribution is already quite flat.  We write down here the values
of $\rho_q(\xi)$ at $\xi=0.2$ and $\xi=0.5$:
\begin{equation}
\begin{split} 
\rho_q(0.2) = 1.07855849250 \\
\rho_q(0.5) = 1.07855849256  
\end{split}
\end{equation}
Moreover, one can show, from known properties of the Jacobi theta function, 
that the function $\rho_q(\xi)$ is monotonically increasing between
$\xi=0$ and $\xi=1/2$ (and of course it is symmetric around $\xi=1/2$).
Thus it seems that the Boltzmann probabilities in Eq.~(\ref{pBs}) weight the sinusoidal functions 
$|\psi_n(\xi)|^2$ in just the right way as to produce a distribution that, over most of the interval,
is almost as flat as the corresponding classical distribution. 


We now ask whether one can {\em deduce} the Boltzmann weights from a requirement
that the position distribution be very flat.  To address this question, we consider a general 
weighting of the squared energy eigenfunctions:
\begin{equation} \label{infinitesum}
\rho(\xi) = \sum_{n=1}^{\infty} P_n |\psi_n(\xi)|^2,
\end{equation}
where the probabilities $P_n$ are to be determined.  We could ask what values of $P_n$
make $\rho(\xi)$ as flat as possible.  But it is clear that this is not the right question: making
the distribution as flat as possible is like taking the infinite-temperature limit, which is not what
we want.  (The quantum distribution becomes completely flat at infinite temperature, but in that
limit every $P_n$ approaches zero.)  Rather, we want somehow 
to limit the effective temperature so that we can compare the $P_n$'s that emerge from a flatness
condition to the $P^B_n$'s given by the Boltzmann distribution.  

Here we take a fairly crude approach to limiting the effective temperature: we simply restrict the
sum in Eq.~(\ref{infinitesum}) to the first $N$ eigenstates.
\begin{equation}
\rho(\xi) = \sum_{n=1}^{N} P_n |\psi_n(\xi)|^2.
\end{equation}
Now we take as our flatness condition the requirement that the first $2N - 1$ derivatives
of $\rho(\xi)$, evaluated at $\xi = 1/2$ (that is, at the middle of the box), be zero.  We have chosen
the number $2N -1$ so that the number of conditions we are imposing equals the number
of variables.  The odd-numbered derivatives are automatically zero at $\xi = 1/2$ because of the symmetry
of the functions $|\psi_n(\xi)|^2$.  Thus our flatness condition, together with the normalization
condition, yields exactly $N$ linear equations for the $N$ variables $P_n$.  Upon taking the 
derivatives, one finds that the equations are
\begin{equation} \label{equations}
\begin{split}
&\sum_{n=1}^N (-1)^n n^{2m}P_n = 0, \hspace{1cm} m = 1, 2, \ldots, N-1 \\
&\sum_{n=1}^N P_n = 1.
\end{split}
\end{equation}

This system of linear equations can be solved exactly.  The unique solution (see Appendix A) is
\begin{equation} \label{solution}
P_n = A \frac{(2N)!}{(N+n)!(N-n)!},
\end{equation}
where the proportionality constant $A$ is determined by normalization.  This constant comes out
to be
\begin{equation}  \label{A}
A = 2\left[ 2^{2N} - \frac{(2N)!}{N!^2}\right]^{-1}.
\end{equation}
Thus the position distribution $\rho(\xi)$ is made optimally flat, in our sense, by choosing the $P_n$'s to form
essentially half of a binomial distribution.  

\begin{figure}[h]
\begin{center}
\includegraphics[scale=0.9]{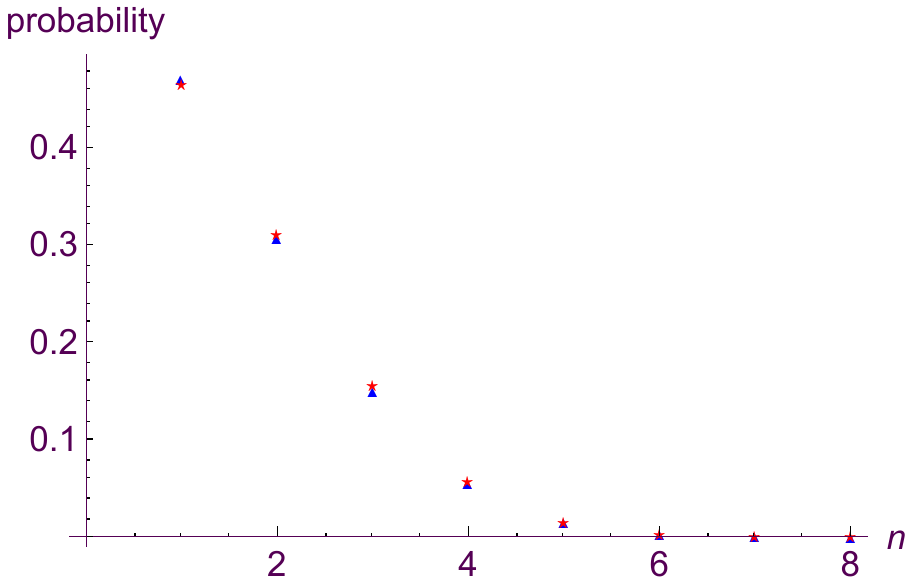}
\end{center}
%
%
\caption{The Boltzmann probabilities (blue triangles) and the probabilities that make
$\rho(\xi)$ optimally flat (red stars).  In this plot $\tau = N = 7$. For much larger values
of $\tau$, it is difficult to see any difference between the two distributions in a plot like this.}
\label{flatBoltz}       
\end{figure}

\begin{figure}[h]
\begin{center}
\includegraphics[scale=0.9]{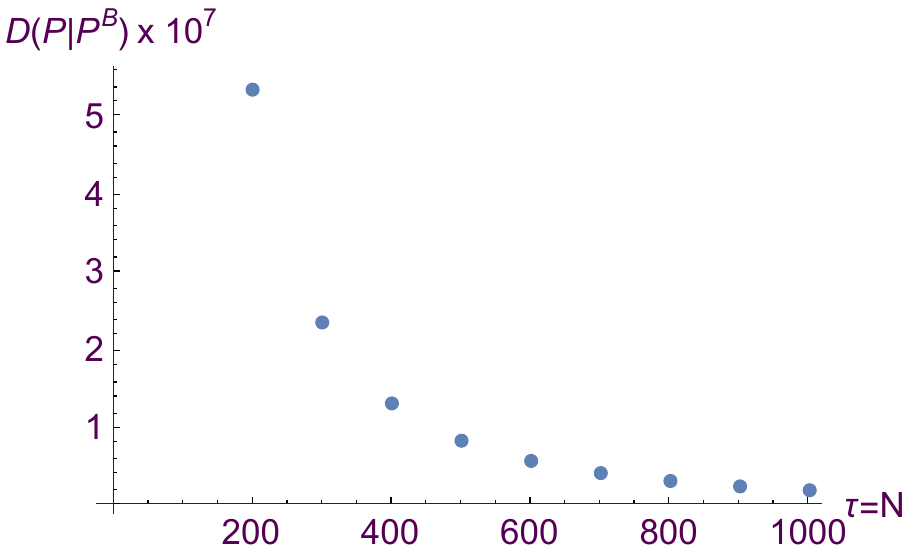}
\end{center}
%
%
\caption{The Kullback-Leibler divergence $D(P|P^B)$ between the flatness-optimizing
distribution $\{P_n\}$ and the Boltzmann distribution $\{P^B_n\}$.  To make the comparison, we
have set the value of $N$ in $P_n$ equal to the value of $\tau$ in $P^B_n$.}
\label{KL1}       
\end{figure}

Now, the Boltzmann formula for a particle in a box, Eq.~(\ref{pBs}), has 
the form of a discrete Gaussian function of $n$.  We know that a binomial distribution ultimately
approaches a Gaussian, so we expect our binomial ``flatness-optimizing" distribution
to look much like the Boltzmann distribution when $N$ is large, as long as we make a judicious
choice of the correspondence between the temperature $\tau$ (for the Boltzmann distribution) and the value of $N$ (for the flatness-optimizing distribution).  It turns out that the choice $\tau = N$
works well.  Fig.~\ref{flatBoltz} shows the two probability distributions for $\tau = N = 7$.  As
the temperature increases, the
flatness-optimizing distribution $\{P_n\}$ approaches the Boltzmann distribution $\{P^B_n\}$ in the sense that the
Kullback-Leibler divergence,
\begin{equation} \label{KL}
D(P|P^B) = \sum_{n=1}^N P_n \ln\left(P_n/P^B_n\right),
\end{equation}
approaches zero, as illustrated in Fig.~\ref{KL1}.  One can in fact show analytically (Appendix B) that this divergence
diminishes as $1/N^2$ for large $N$.  Thus although our flatness-optimizing distribution is
not identical to the Boltzmann distribution---it could not be, since it is limited to the first
$N$ eigenstates---it approaches the Boltzmann distribution in a rather strong sense
as the temperature increases.  We regard this asymptotic
result as interesting, because it seems so different from the 
usual way of deriving these probabilities.  

Regarding the infinite-temperature limit, we note that merely requiring 
the position distribution $\rho(\xi)$ to become 
flat in this limit is
not enough to see anything special about the Boltzmann probabilities $P^B_n$.  Consider,
for example, the
alternative set of probabilities $P^A_n =$ \hbox{$(e^{1/\tau} - 1) e^{-n/\tau}$}.  (The ``A'' is for ``alternative.'')  The resulting position distribution,
\begin{equation}
\rho^A(\xi) = \sum_{n=1}^{\infty} P^A_n |\psi_n(\xi)|^2 = 
\frac{(e^{1/\tau}+1)(1-\cos (2 \pi \xi))}{e^{1/\tau}-2\cos (2 \pi \xi) + e^{-1/\tau}}\; \, ,
\end{equation}
likewise becomes arbitrarily flat as $\tau$ goes to infinity.  But $P^A_n$ is clearly very different from
$P^B_n$.  Evidently one needs some kind of finite-temperature flatness condition in order 
to arrive at something that
approaches the Boltzmann distribution in the infinite-temperature limit.  

Again we emphasize that, in our flatness-optimizing problem, we have not put into the mathematics
the energy eigenvalues.  
(We have put into the mathematics the squared eigenfunctions $|\psi_n(\xi)|^2$, together with an {\em ordering} of
these eigenfunctions, but we have
not explicitly 
put in the associated energies.)  So, to the extent that the Boltzmann formula emerges from
the flatness condition, part of what emerges is the fact that the exponent in this formula is proportional
to $n^2$.  

Our insistence on flatness of the position distribution may bring to mind Jaynes' characterization
of the thermal state as the state of maximum entropy for a fixed value of the average energy \cite{Jaynes}.
Perhaps even more relevant is the recent work by Anz\`a and Vedral on states for which the 
Shannon entropy of a particular {\em observable} is maximized 
for fixed average energy \cite{Anza}.
One might suppose from the above results that, for the particle in a box, the {\em position} is an
observable whose differential 
entropy $h = -\int_0^1 \rho(\xi) \ln \rho(\xi) d\xi$ is maximized in the thermal state; that is, one might imagine that the observed flatness
of the function $\rho(\xi)$ is a manifestation of position-entropy maximization.  
However, it turns out that this is not the case, at least not if the average energy is being held fixed.  It is easy to find
states with the same average energy as the thermal state but exhibiting a greater 
position-entropy.  

Of course, in our problem we did not hold the average energy fixed.  Rather, we fixed
the value of $N$, which labels the highest-energy eigenstate allowed to have nonzero probability.  
But if one tries maximizing the position-entropy $h$ while holding $N$ fixed,
one finds that the result is quite different from the Boltzmann distribution.  
So, if there is a connection between
our observations and Jaynes' principle, it must be a subtler connection.

\section{Harmonic oscillator}
\label{sec:3}
Consider a one-dimensional harmonic oscillator in thermal equilibrium.  The potential 
energy function is $V(x) = (1/2)m\omega^2x^2$, where $m$ is the mass and $\omega$ is the
angular frequency.  In terms of a rescaled position variable $\xi = x\sqrt{m\omega/\hbar}$ and 
a rescaled temperature $\tau = kT/(\hbar\omega)$, the squared energy eigenfunctions
are
\begin{equation} \label{oscpsi2}
|\psi_n(\xi)|^2 = \frac{1}{2^n n! \sqrt{\pi}}e^{-\xi^2}H^2_n(\xi), \hspace{3mm} n=0, 1, \ldots,
\end{equation}
and the Boltzmann probabilities are
\begin{equation} \label{hopBs}
P^B_n = (1-e^{-1/\tau} )e^{-n/\tau}.
\end{equation}
In Eq.~(\ref{oscpsi2}), $H_n(\xi)$ is the Hermite polynomial
\begin{equation}
H_n(\xi) = (-1)^n e^{\xi^2} \frac{d^n}{d\xi^n}e^{-\xi^2}.
\end{equation}
The quantum mechanical thermal position distribution is therefore
\begin{equation} \label{hoqrho}
\begin{split}
\rho_q(\xi) &= \sum_{n=0}^\infty P^B_n |\psi_n(\xi)|^2 \\
&= \frac{1}{\sqrt{\pi}} (1-e^{-1/\tau} )e^{-\xi^2}\sum_{n=0}^\infty \frac{e^{-n/\tau}}{2^n n!}H^2_n(\xi).
\end{split}
\end{equation}
It is well known that the sum in Eq.~(\ref{hoqrho}) comes out to be an exact Gaussian.  
The distribution $\rho_q(\xi)$ can be rewritten
as \cite{Hillery}
\begin{equation}  \label{hoq}
\rho_q(\xi) = \frac{1}{\sqrt{2 \pi \langle \xi^2\rangle}}e^{-\frac{\xi^2}{2\langle \xi^2\rangle}},
\end{equation}
where 
\begin{equation}  \label{avexi2}
\langle \xi^2 \rangle = \frac{1}{2 \tanh(\frac{1}{2\tau})}.
\end{equation}
Meanwhile the classical position distribution, proportional to $e^{-V(x)/kT}$, when written 
in terms of our rescaled variables becomes
\begin{equation}  \label{hoc}
\rho_c(\xi) = \frac{1}{\sqrt{2 \pi \tau}} e^{-\frac{\xi^2}{2\tau}}.
\end{equation}
From Eqs.~(\ref{hoq}) and (\ref{hoc}) it is clear that $\rho_q$ differs from $\rho_c$ only because
$\langle \xi^2 \rangle$ is not the same as $\tau$ (reflecting the quantum violation of the 
classical equipartition theorem).

In the preceding section, we asked what probability distribution $\{P_n\}$ would make the 
quantum position distribution for the particle in a box 
as flat as possible (in a certain sense), so that it would be
as close as possible to the uniform classical distribution.  We found that the resulting probability distribution $\{P_n\}$, while
not equal to the Boltzmann distribution $\{P^B_n\}$, did approach that distribution in the high-temperature
limit.   We can ask a similar question here.
Given the classical position distribution $\rho_c(\xi)$ for a given value of $\tau$, and 
given the squared energy eigenstates
$|\psi_n(\xi)|^2$ of the harmonic oscillator, we ask how the probabilities $P_n$ should be chosen so that
the weighted average,
\begin{equation}  \label{horho}
\rho(\xi) = \sum_{n=0}^{\infty} P_n |\psi_n(\xi)|^2 ,
\end{equation}
makes the ratio $r(\xi) = \rho(\xi)/\rho_c(\xi)$ as flat as possible.
Do the optimal 
probabilities $P_n$ look like the Boltzmann probabilities given in Eq.~(\ref{hopBs}), and do they
approach those probabilities at high temperature?

In this case we can take ``as flat as possible'' to mean ``exactly constant": 
as long as $\tau$ is at least 1/2, we can simply
choose the $P_n$'s to be the {\em quantum} probabilities evaluated at whatever temperature
is required to make the spread of the quantum Gaussian equal to that of the classical Gaussian.  
Then $\rho(\xi)$ of Eq.~(\ref{horho}) will be exactly equal to the classical distribution
$\rho_c(\xi)$.  
(If $\tau$ is less than 1/2, the classical spread is less than the spread of the quantum ground state.
A quantum particle cannot achieve such a narrow distribution.)  
Specifically, we choose $P_n$ to be
\begin{equation}  \label{aaa}
P_n = (1 - e^{-1/\tau_e})e^{-n/\tau_e},
\end{equation}
where the effective temperature $\tau_e$ is given by
\begin{equation}  \label{taue}
\frac{1}{2\tau_e} = \tanh^{-1}\left(\frac{1}{2 \tau}\right).
\end{equation}
That is, $\tau_e$ is the value we have to substitute for the $\tau$ in Eq.~(\ref{avexi2})
in order to make $\langle \xi^2 \rangle$---the quantum variance---equal to the actual value of $\tau$,
which is the classical variance.
Equations (\ref{aaa}) and (\ref{taue}) define our optimal $P_n$'s for the given value of $\tau$.
(The latter equation makes clear that we cannot achieve equality with the classical distribution
if $\tau$ is less than 1/2.)

As we did in the preceding section, we again ask whether the probabilities $P_n$ that
make the weighted average closest to the classical distribution approach the Boltzmann
distribution $P^B_n$ in the Kullback-Leibler sense as the temperature approaches infinity.  
In this case the Kullback-Leibler divergence can be worked out exactly.  One finds that
\begin{equation}
D(P|P^B) = \log \left( \frac{1- e^{-1/\tau_e}}{1 - e^{-1/\tau}}\right) + \frac{e^{-1/\tau_e}}{1-e^{-1/\tau_e}}
\left( \frac{1}{\tau} - \frac{1}{\tau_e} \right).
\end{equation}
Expanding this function out in powers of $1/\tau$ (using Eq.~(\ref{taue})), we get that 
\begin{equation}
D(P|P^B) = \frac{1}{288}\cdot \frac{1}{\tau^4} + \, {\mathcal O}\left(\frac{1}{\tau^6}\right).
\end{equation}
So the ``classical-mimicking" distribution $\{P_n\}$ does approach the Boltzmann distribution
rapidly.  

In a certain sense, our results for the harmonic oscillator are not as impressive as for
the particle in a box.  In order to get our ``classical-mimicking'' probabilities for the 
harmonic oscillator, we needed to invoke, at the outset, the classical Boltzmann factor
$e^{-V(x)/kT}$.  Thus it may not be terribly surprising that the quantum Boltzmann factor,
$e^{-E_n/kT}$, emerged from the calculation (asymptotically).  
(In Section VI, we ask whether what we have seen in this example
points to a special property of the exponential form of the Boltzmann factor.)
Still, though, there is the notable fact that we did not
put the energy eigenvalues into our calculation.  
They are part of what emerges.

\section{Particle in a linear potential}
\label{sec:4}
We now consider a case that combines features of both of the previous examples: a hard wall 
and an otherwise continuously varying potential.  For a particle of mass $m$ in 
one dimension, let the potential energy
function be $V(x) = \alpha x$ for $x \ge 0$, with a hard wall at $x=0$.  We use the rescaled
position and temperature variables
\begin{equation}
\xi = x\left(\frac{2 m \alpha}{\hbar^2}\right)^{1/3}   \hspace{1cm}
\tau = kT \left(\frac{2 m}{\hbar^2 \alpha^2}\right)^{1/3}.
\end{equation}
Then the squared energy eigenfunctions are (for $\xi \ge 0 $)
\begin{equation}
|\psi_n(\xi)|^2 = \frac{\Ai(\xi + u_n)^2}{\Ai'(u_n)^2}.
\end{equation}
Here $\Ai$ is the Airy function, $\Ai'$ is its derivative, and $u_n$ is the $n$th zero 
of the Airy function (all the $u_n$'s are negative).  The Boltzmann probabilities with which
the $|\psi_n|^2$'s are weighted in thermal equilibrium are 
\begin{equation} \label{BoltzmannP}
P^B_n = \frac{1}{Z} e^{u_n/\tau},
\end{equation}
where $Z = \sum_{n=1}^\infty e^{u_n/\tau}$.  So the quantum mechanical position distribution
is 
\begin{equation}
\rho_q(\xi) = \sum_{n=1}^\infty P^B_n |\psi_n(\xi)|^2 
= \frac{1}{Z} \sum_{n=1}^\infty \frac{e^{u_n/\tau}\Ai(\xi + u_n)^2}{\Ai'(u_n)^2}
\end{equation}
The classical position distribution is simply
\begin{equation}
\rho_c(\xi) = (1/\tau)e^{-\xi/\tau}.
\end{equation}

\begin{figure}[h]
\begin{center}
\includegraphics[scale=0.9]{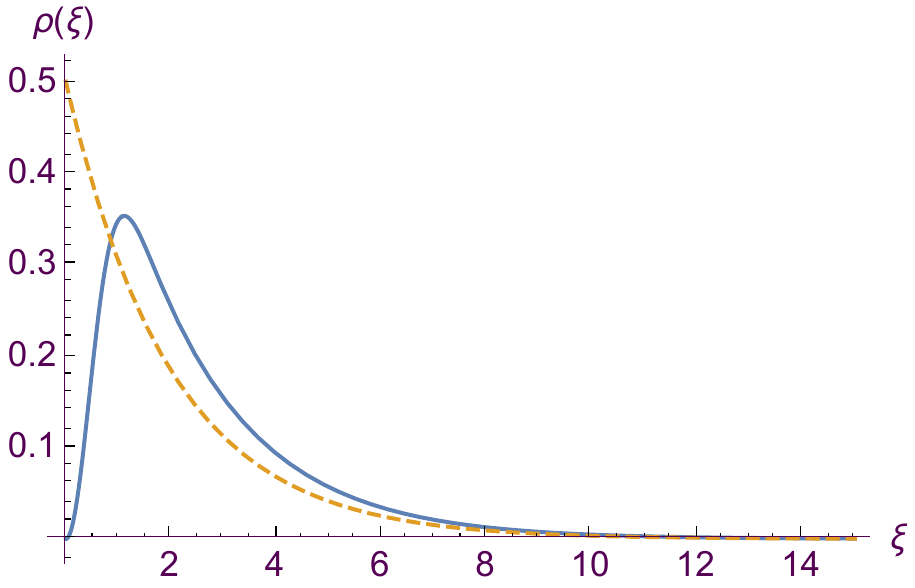}
\end{center}
%
%
\caption{The quantum (solid blue) and classical (dashed gold) position distributions for a 
particle in a linear potential with a hard wall at $\xi = 0$.  Here $\tau = 2$.}
\label{linqc}       
\end{figure}

\begin{figure}[h]
\begin{center}
\includegraphics[scale=0.9]{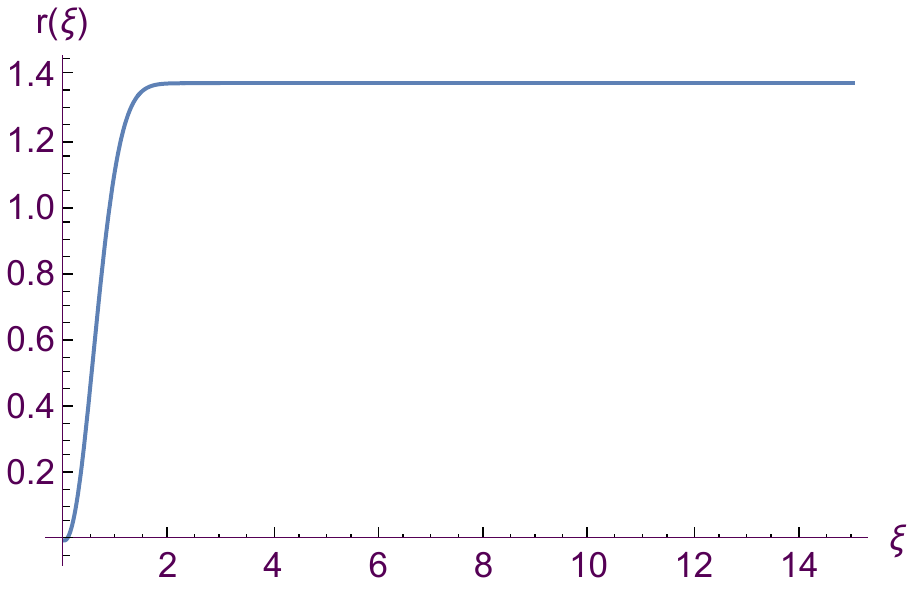}
\end{center}
%
%
\caption{The ratio $r(\xi) = \rho_q(\xi)/\rho_c(\xi)$ for $\tau = 2$.  Beyond $\xi=6$, this ratio appears to be
constant to 30 decimal places (though not exactly constant).}
\label{linratio}       
\end{figure}

\begin{figure}[h]
\begin{center}
\includegraphics[scale=0.9]{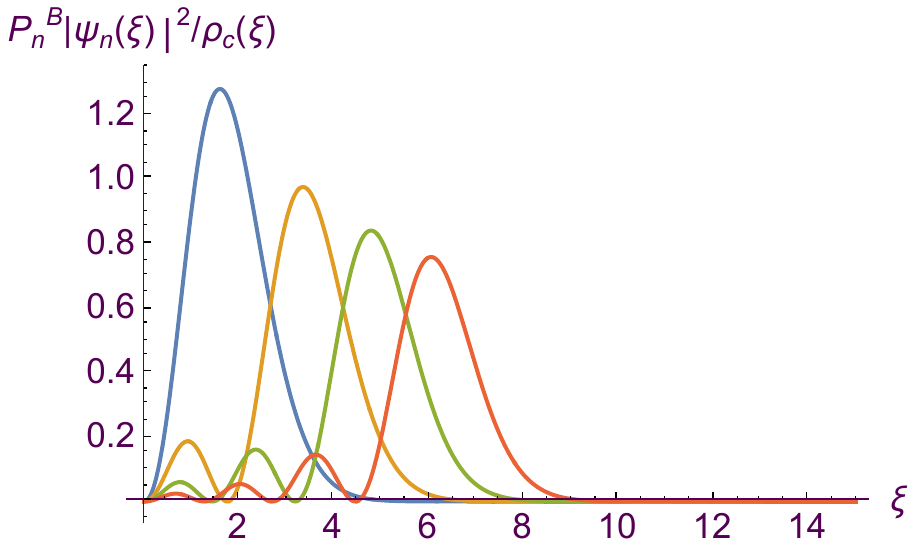}
\end{center}
%
%
\caption{The first four squared energy eigenfunctions for the linear potential, each weighted
by the appropriate Boltzmann weight $P^B_n$ and divided by $\rho_c(\xi)$.  The sum of these curves and their successors
yields the function $r(\xi)$ of Fig.~\ref{linratio}.}
\label{linfirstfour}       
\end{figure}

Fig.~\ref{linqc} shows both of these position distributions for $\tau=2$, and Fig.~\ref{linratio} shows
the ratio $r(\xi) = \rho_q(\xi)/\rho_c(\xi)$.  At this relatively 
low value of the temperature, the two distributions 
are notably different, but the ratio is remarkably uniform beyond a narrow boundary region near 
the wall.  For $\xi > 6$, the ratio appears to be constant to 30 decimal places (though not 
exactly constant).  Fig.~\ref{linfirstfour} shows the contributions to 
the ratio $r(\xi)$ from the first four energy eigenfunctions.  Note that the Boltzmann weights
seem to 
give these contributions precisely the right height so that they sum to an almost constant
function (except close to the wall).  

For this example, we have not been able to formulate a sharp and tractable optimization problem
that would define a set of $P_n$'s making the ratio
\begin{equation}
r(\xi) = \frac{\rho(\xi)}{\rho_c(\xi)} = \frac{\sum_{n=1}^\infty P_n |\psi_n(\xi)|^2}{\rho_c(\xi)}
\end{equation}
``as flat as possible.''  (But see Section V, in which we propose one reasonable 
formulation of the problem, though not a formulation that we have been able to solve.)  It is nevertheless possible to interpret ``as flat as
possible'' in a weaker sense, which does at least seem to pick out the Boltzmann values $P^B_n$ for large values of $n$.  We can see this through the following informal argument.

We begin by writing $r(\xi)$ explicitly in terms of the Airy function:
\begin{equation}  \label{rr}
r(\xi) = 
\tau e^{\xi/\tau} \sum_{n=1}^{\infty} P_n \frac{\Ai(\xi + u_n)^2}{\Ai'(u_n)^2}.
\end{equation}
Let us now define $R_n$ by
\begin{equation} \label{Rdef}
P_n = \Ai'(u_n)^2 e^{u_n/\tau} R_n.
\end{equation}
In terms of $R_n$, we have
\begin{equation} \label{rrrr}
r(\xi) = \tau \sum_{n=1}^\infty R_n f(\xi + u_n),
\end{equation}
where $f(z) = e^{z/\tau} \Ai(z)^2$.  That is, $r(\xi)$ (for $\xi \ge 0$) is a weighted sum of an infinite number
of shifted versions of the {\em same} function $f(z)$.  

Now, if $r(\xi)$ is to be almost constant for large $\xi$, in particular we must have that, for 
large $s$ and $t$ with $t > s$, 
\begin{equation}
\int_{|u_s|}^{|u_t|} r(\xi) d\xi \approx A(\tau)(|u_t| - |u_s|),
\end{equation}
where $A(\tau)$ is independent of $s$ and $t$.  We now substitute the expression in
Eq.~(\ref{rrrr}) for the $r(\xi)$ in this integral. Our condition for flatness becomes
\begin{equation}  \label{sumsum}
\tau \sum_{n=1}^\infty R_n \int_{|u_s|}^{|u_t|} f(\xi + u_n) d\xi \approx A(\tau)(|u_t| - |u_s|).
\end{equation}

The function $f(z)$ has a characteristic width that depends on the value of $\tau$.  
(If we normalize $f(z)$ so as to turn it into a probability distribution, we find 
that the width of $f(z)$, quantified by the standard deviation $\Updelta z$, is approximately equal 
to $\tau/\sqrt{2}$ for large $\tau$.)  Let us now
consider the special case where both $|u_t| - |u_s|$ and $|u_s|$ are much larger than this characteristic width.  
Then for most of those terms in the sum in Eq.~(\ref{sumsum}) for which $s < n < t$, the range of integration completely spans the 
region where the integrand is significantly different from zero, and we can write
\begin{equation}
\int_{|u_s|}^{|u_t|} f(\xi + u_n) d\xi \approx \int_{-\infty}^{\infty} f(z) dz = \frac{\sqrt{\tau}}{2\sqrt{\pi}}
e^{1/(12 \tau^3)}.
\end{equation}
Meanwhile, for most of those terms for which $n$ is outside of the range \hbox{$s < n < t$}, the 
integral is approximately zero.  We can therefore further approximate our condition for flatness (Eq.~(\ref{sumsum})) to be
\begin{equation}  \label{t}
\sum_{n=s}^t R_n \approx B(\tau)(|u_t| - |u_s|),
\end{equation}
where $B(\tau)$ is another function of $\tau$ that does not depend on $s$ and $t$.  
For Eq.~(\ref{t}) to be true regardless of the values of $s$ and $t$, we want $R_n$ to have the form
\begin{equation}
R_{n} \approx B(\tau)(|u_n| - |u_{n-1}|).
\end{equation}
(Or we could say, $R_{n} \approx B(\tau)(|u_{n+1}| - |u_{n}|)$.  At this level of approximation,
the two expressions are equivalent.)
It is known that $|u_n| - |u_{n-1}|$ is roughly proportional to $n^{-1/3}$ for large $n$ \cite{NIST}.  So
we want $R_n$ to be roughly proportional to $n^{-1/3}$, and, according to Eq.~(\ref{Rdef}), we
therefore want $P_n$ to satisfy
\begin{equation}
P_n \propto n^{-1/3}\Ai'(u_n)^2 e^{u_n/\tau}.
\end{equation}
Finally, it is also known that for large $n$, the combination $n^{-1/3}\Ai'(u_n)^2$ is nearly 
constant.  (The limiting value for infinite $n$ is $(3/(2 \pi^2) )^{1/3}$ \cite{NIST}.)  
We therefore arrive at
\begin{equation} \label{finalP}
P_n \propto e^{u_n/\tau},
\end{equation}
where we expect the proportionality constant to depend on the temperature.  But 
Eq.~(\ref{finalP}) agrees with the form of the Boltzmann probabilities (Eq.~(\ref{BoltzmannP})).
In this way we can see how, at least for large $n$, 
the Boltzmann probabilities $P^B_n \propto e^{u_n/\tau}$ naturally arise if one is trying to 
make the ratio $r(\xi)$ roughly constant.  

Note that, as in the 
earlier sections, our argument is based only on the 
squared energy eigenfunctions $|\psi_n(\xi)|^2$ and not the associated energies.  (We may not
be able to
{\em write down} those eigenfunctions without knowing the energies, but conceptually it is still the squared
eigenfunctions and not the energies that enter the argument.)

\section{A general definition of ``as flat as possible''?}
\label{sec:5}

The main problem we have considered is this: Given the squares of the energy eigenfunctions
in the position representation,
how can these eigenfunctions be weighted so as to yield a function $\rho(x)$ for which
$\rho(x)/\rho_c(x)$ is ``as flat as possible"?  (Again, $\rho_c(x)$ is the classical thermal position distribution at a given temperature.)  In the first two examples (the particle in a box and the harmonic oscillator), the probability distributions
generated in this way turned out to approach, in the Kullback-Leibler sense, the quantum Boltzmann distribution in the limit of infinite temperature.  The third example (the particle in a linear
potential) also seemed to produce something like the Boltzmann distribution, though our
argument in that case is not rigorous.  

But so far we have interpreted ``as flat as possible'' in an {\em ad hoc} way.  It is interesting to ask whether one can come up with a general definition of ``as flat as possible'' that 
would yield the Boltzmann probabilities, in this asymptotic sense, for a large class of systems.  
Our results for the three examples we have
considered in this paper suggest
that the following approach might work.  

For now let us restrict ourselves to the one-dimensional case.  Let the position $x$ run from $a$ to $b$, where $a$ might be $-\infty$ and $b$ might be $\infty$.  Given the classical position 
distribution $\rho_c(x)$ associated with a given temperature, let us begin by defining a new 
variable $y$ by
\begin{equation}
y = \int_{a}^x \rho_c(x') dx'.
\end{equation}
Then $y$ runs from $0$ to $1$ (since $\rho_c$ is normalized), and the classical distribution for $y$ is the uniform distribution (just as it is for $\xi$ in the case of the particle in a box).  We now consider, for each energy eigenstate
$\psi_n(x)$, the corresponding distribution of the variable $y$---let us 
call it $\sigma_n(y)$---so that a weighted average of these $\sigma_n(y)$'s would
yield some overall distribution $\sigma(y)$.   We want $\sigma(y)$ to be ``as flat as possible.''  
Before we write down our proposed definition of ``as flat as possible,'' let us 
illustrate the construction of $\sigma(y)$
by considering the example of the linear potential.  

\begin{figure}[h]
\begin{center}
\includegraphics[scale=0.9]{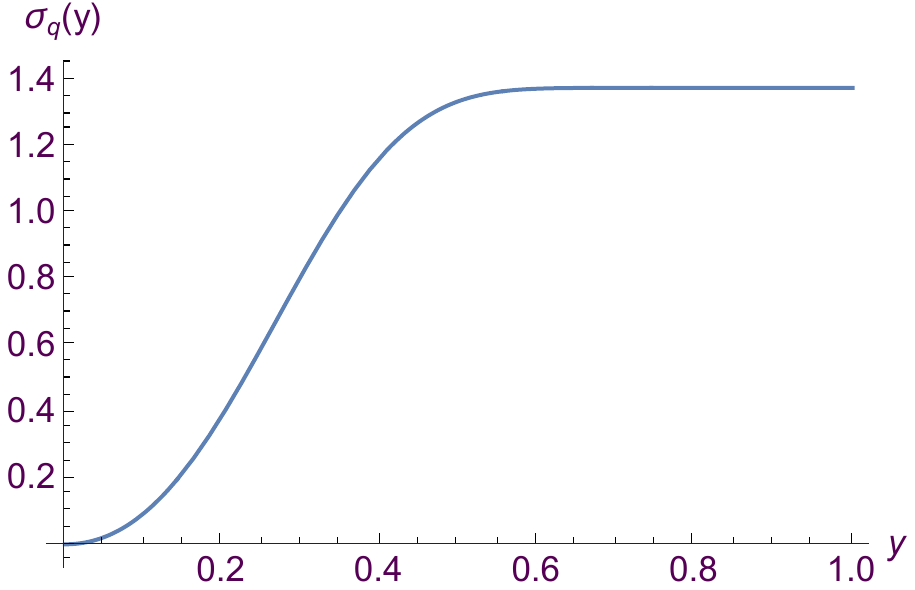}
\end{center}
%
%
\caption{The quantum position distribution for a particle in a linear potential 
when position is parameterized by the variable $y$.  Here $\tau = 2$.}
\label{liny}       
\end{figure}

Recall that for the linear potential, in terms of the dimensionless variable $\xi$, we have
\begin{equation}
\rho_c(\xi) = (1/\tau)e^{-\xi/\tau}.
\end{equation}
So 
\begin{equation}
y = \int_{0}^\xi \rho_c(\xi') d\xi' = 1 - e^{-\xi/\tau}.
\end{equation}
Let $\rho_n(\xi) = |\psi_n(\xi)|^2 = \Ai(\xi+u_n)^2/\Ai'(u_n)^2$ be the $n$th squared eigenfunction.
The corresponding distribution $\sigma_n(y)$ is defined by
\begin{equation}
\sigma_n(y) dy = \rho_n(\xi)d\xi.
\end{equation}
So in this example,
\begin{equation}
\sigma_n(y) = \rho_n(\xi)\frac{d\xi}{dy} = \left( \frac{\tau}{1-y} \right)\frac{\Ai[-\tau \ln(1-y) + u_n]^2}{\Ai'(u_n)^2}.
\end{equation}
A general weighted average would be 
\begin{equation}
\sigma(y) = \sum_{n=1}^\infty P_n \sigma_n(y).
\end{equation}
For the special case where the $P_n$'s are the Boltzmann weights $P_n^B \propto e^{u_n/\tau}$,
for $\tau = 2$ 
we get the curve $\sigma_q(y)$ shown in Fig.~\ref{liny}.  

We now present our proposed definition of ``as flat as possible.'' For any small $\epsilon$
greater than zero, let $Y_\epsilon$ be the set of all values of $y$ for which
$\sigma(y) \ge \sigma_{max} - \epsilon$, where $\sigma_{max}$ is the maximum value of $\sigma(y)$,
if the maximum exists.  (For the harmonic oscillator with $\tau < 1/2$, the function $\sigma(y)$ will 
take arbitrarily large values no matter how the eigenstates are weighted, because 
each $\sigma_n(y)$ itself takes arbitrarily large values.  So our definition will not apply to that
case.)
For each value of $\epsilon$, we can look for a distribution $\{P^\epsilon_n\}$ that maximizes the {\em measure} of 
the set $Y_\epsilon$ (for the standard measure on the interval $[0,1]$).  (For an exceptional case such as the particle in a box, 
in which the classical position distribution is independent of temperature, we need to limit
the effective temperature in some way in order to get a solution, as we did in Section II.)
We can then define our optimal distribution $\{P_n\}$ to be the limit
of $\{P^\epsilon_n\}$ as $\epsilon$ approaches zero, if the limit exists.  This formulation of the 
problem is consistent with our definition of ``as flat as possible'' for the harmonic oscillator
(for $\tau \ge 1/2$).
It is also plausibly consistent with what we did for the particle in a box: if the sole maximum
of $r(\xi)$ occurs at the center of the box, then one wants the derivatives of as many orders
as possible to be zero at the center in order to make $r(\xi)$ as constant as possible
near that location.  For the third example, we have nothing rigorous to report, 
but this general formulation seems sensible
for that case as well. 

We can express this same proposed definition in terms of the original variable $x$.
For any small $\epsilon$ greater than zero, let $X_\epsilon$ be the set of all values of
$x$ for which $r(x) \ge r_{max} - \epsilon$, where, as usual, $r(x) = \rho(x)/\rho_c(x)$
and $\rho(x) = \sum_n P_n |\psi_n(x)|^2$.  Here $r_{max}$ is the supremum
of $r(x)$ over the range of $x$ (if the supremum exists).  
Now we look for a distribution $\{P^\epsilon_n\}$
for which the {\em probability} of the set $X_\epsilon$ is greatest, where the probability
is computed using the classical thermal distribution $\rho_c(x)$.  That is, the probability 
we want to maximize is 
\begin{equation}
\hbox{probability of $X_\epsilon$} \; = \int_{x \in X_\epsilon} \rho_c(x) dx.
\end{equation}
We then take the
limit of $\{P^\epsilon_n\}$ as $\epsilon$ goes to zero.  

In a higher-dimensional configuration space, there is an obvious generalization of the
above criterion.  In fact, the preceding paragraph can simply be reinterpreted, with
$x$ referring to the location in a multidimensional space.  

We also need to consider the possibility of a degeneracy in energy.  Within a degenerate subspace, there is no unique set of squared energy eigenfunctions, since
 one has a choice of basis for the subspace.  In that case, we can simply
use, for each degenerate subspace, the configuration-space distribution associated with the completely mixed
state over that subspace.  The question would then be how to weight the distributions
associated with distinct energies so as to make $r(x)$ as flat as possible.  

Now, there certainly exist systems for which our approach does not work.  A simple example
is a particle on a one-dimensional ring (that is, a one-dimensional box with periodic boundary
conditions).  In that case every energy eigensubspace is associated with a uniform 
position distribution.  So the quantum position distribution automatically agrees with the
classical position distribution regardless of how these subspaces are weighted relative
to each other.  Moreover, it is not clear that using the other dimension of phase space would
help in this case: in the quantum case the momentum does not even take the same set of
values as in the classical case.  Perhaps it should not be surprising if our approach, based 
largely on 
a comparison between classical and quantum physics, does not apply to a system
for which the classical and quantum phase spaces are not the same \cite{Gonzales,Chadzitaskos}.

\section{Discussion}
\label{sec:6}

Normally, what one needs in order to write down the Boltzmann probabilities are the 
energy eigenvalues.  Here we have not used the energy eigenvalues; the inputs to our 
calculation are instead the squares of the energy eigenfunctions, together with the classical
thermal
position distribution we are trying to match.  Now, one might argue that once we are given the
classical position distribution, we can use it,
together with the Schr\"odinger equation, to find the energy eigenvalues, since the 
potential energy function can be extracted from the 
classical distribution (up to an additive constant).  The
energy eigenvalues are in this sense implicitly present in our inputs.  However, we 
have not put the full Schr\"odinger equation into our mathematics.  So it is still
intriguing that the
optimal weights somehow know that they are supposed to approach the Boltzmann
weights.  

One might also argue that our results are merely a consequence of the fact that, even for a single energy eigenstate, there is typically a close
correspondence between the quantum position distribution and a classically defined
probability distribution associated with the given energy \cite{Robinett,Doncheski,Bernal,Semay}.  
However, this fact cannot fully explain our observations.  The oscillating 
quantum probability distributions associated with the energy eigenstates, 
when combined in a weighted 
average, are capable of producing a wide range of functions.  The weights in the average
need to be chosen in just the right way if they are to produce a position distribution similar to the 
classical distribution.  In our examples, the Boltzmann weights seem to do this.


It is important to recognize that our three examples should be placed into two distinct
categories.  One category is represented by our last two examples, 
the harmonic oscillator and the linear potential.  In those cases,
the classical Boltzmann distribution was a crucial piece of the input, and
the fact that something like the quantum Boltzmann distribution emerged from 
the calculation surely reflects the fact that the quantum and classical Boltzmann factors
exhibit the same dependence on energy, and that ``energy'' has the same physical meaning in both
contexts.

In contrast, for the case of the particle in a box (representing the other category), the 
classical Boltzmann factor played
no essential role at all.  The only fact from classical physics we used 
in that case was the uniformity of the thermal position distribution.
This uniformity has nothing particularly to do with the classical Boltzmann factor, $e^{-V(x)/kT}$;
any other function of $V(x)$ would also have given rise to the uniform distribution,
since $V(x)$ itself is uniform.
So, in this one example, the fact that something very close to the
Boltzmann distribution emerged from the calculation does not say much at all about the 
relation between quantum and classical mechanics.  Rather, it seems to suggest a relationship
between quantum mechanics and {\em statistical mechanics}.  We started with the 
squared eigenfunctions, which come from quantum mechanics, and we 
found that the ``flatness-optimizing'' weighting of these eigenfunctions approaches the 
quantum Boltzmann distribution, which belongs to statistical mechanics \cite{fn2}.  

Of course we cannot draw general conclusions from this one example.  It would be 
interesting to carry out a similar analysis for other ``boxes'' in which the potential energy function
is likewise uniform, {\em e.g.}, two-dimensional boxes of various shapes.  To do that, one would 
ideally use
a more general method of limiting the effective temperature than the method we used in Section II.  
Perhaps one could fix the value of the {\em Shannon entropy} of the distribution $\{P_n\}$.  

It is also interesting to ask whether there is an alternative way of framing the problem for our
two ``non-box'' examples---or indeed for any system where the potential energy has a non-trivial 
dependence on the configuration---so as to see whether the particular functional form
$e^{-E/kT}$, applied to both quantum and classical physics, is somehow favored
over other functions of $E/kT$ as the form most conducive to agreement
between the quantum and classical configuration-space distributions (as indeed it does 
seem to be favored
for the particle in the box).  

Here is one way we might imagine framing the problem.  For a given system, imagine
a hypothetical world where the classical thermal distribution over phase space has the form 
$(1/Z)f(E(x,p)/kT)$; here $E(x,p)$ is the 
classical energy function, $f$ is a suitably normalizable but otherwise arbitrary 
non-negative function, and $Z$ is the
integral of $f(E(x,p)/kT)$ over phase space.  The classical configuration-space distribution 
(in this hypothetical world)
is then the integral of $(1/Z)f(E(x,p)/kT)$ over $p$.  Now solve the problem posed in Section V.  
That is, find the weights $\hat{P}_n$ of the quantum energy eigenfunctions that 
optimize the flatness of the ratio $\hat{r}(x)=\hat{\rho}(x)/\hat{\rho}_c(x)$, where $\hat{\rho}_c(x)$ is the 
distribution obtained from $f$.  (We use hats to indicate that we are in the hypothetical 
world where the classical distribution function is defined by $f$.)  Now we ask: For large values of $T$,
does the optimal distribution $\{\hat{P}_n\}$
approach the probability distribution proportional to $f(E_n/kT)$, where the $E_n$'s are the
energy eigenvalues?  If the answer is ``yes" for 
almost any function $f$, then the problem does not demonstrate any special property
of the Boltzmann factor $e^{-E/kT}$.  On the other hand, if the answer is ``yes'' only if $f(z)$ is proportional
to $e^{-z}$, then one can say that the Boltzmann factor is especially conducive to
a quantum-classical correspondence.  Essentially, what we are asking here is whether an arbitrary 
dependence on $E/kT$ would produce results like those we have seen for the Boltzmann distribution
in Sections III and IV.

Of course we already have at least two standard derivations of the Boltzmann distribution.  In addition
to the Jaynes approach mentioned earlier, in which information-theoretic entropy is
maximized, 
there is also the argument found in most textbooks
on statistical mechanics, which starts from the assumption that the system of interest
is embedded in a much larger system described by the microcanonical
ensemble.  Moreover, in recent years much progress has been made toward understanding
why, and under what circumstances, one can expect many of the consequences of the 
microcanonical ensemble to apply even to an isolated system in a pure quantum state \cite{Popescu,Dziarmaga,Polkovnikov,Nandkishore,Eisert,Gogolin,DAlessio}.

If, then, the results we have obtained in this paper can be sharpened and generalized to a large class of systems, it would
seem that the Boltzmann probabilities are, in a sense, overdetermined.  They are determined 
by the standard arguments---entropy maximization or the argument from the microcanonical ensemble---and
they are also determined, at least in an asymptotic sense, by the apparently independent criterion of a close correspondence
between quantum and classical physics.  This overdetermination would suggest that what
appear to be independent arguments are not actually independent, and that there exist underlying connections between
quantum, classical, and statistical mechanics that we have not identified.  Conceivably, what we
are seeing in this paper is another aspect of the relationship, mentioned in the preceding paragraph, between the structure of quantum mechanics
and the foundations of statistical mechanics.

Of course, another possibility is that the observations we have made here are peculiar to our
examples and do not generalize, in which case they should be regarded merely as interesting curiosities without deeper significance.  The next step is therefore to see to what extent these
results can be extended.  We hope to explore this question in future work.

\begin{acknowledgements}
We would like to thank Daniel Aalberts, Fabio Anz\`a, Steve Miller, Ben Schumacher, and Swati Singh for insightful discussions and comments.  
\end{acknowledgements}

\section*{Appendix A}

Here we prove that Eq.~(\ref{solution}) is the unique solution to the system of linear equations
given in Eq.~(\ref{equations}).  First we show that it is a solution; then we show that the solution
is unique.

We begin with the relation
\begin{equation}
\begin{split}
(1 + x)^{2N} &= \sum_{l=0}^{2N} \frac{(2N)!}{l! (2N - l)!} x^l \\
&= x^N \sum_{n=-N}^N \frac{(2N)!}{(N+n)!(N-n)!}x^n,
\end{split}
\end{equation}
where, in the last expression, we have replaced the summation variable $l$ with $n = l - N$.
Dividing by $x^N$, we get
\begin{equation}
\left( 2 + x + \frac{1}{x} \right)^N = \sum_{n=-N}^N \frac{(2N)!}{(N+n)!(N-n)!}x^n.
\end{equation}
Now apply the operation $(x \frac{d}{dx})^{2m}$ to both sides:
\begin{equation}
\left( x \frac{d}{dx} \right)^{2m} \left( 2 + x + \frac{1}{x} \right)^N
= \sum_{n=-N}^N \frac{(2N)!}{(N+n)!(N-n)!} n^{2m} x^n.
\end{equation}
Evaluating both sides at $x = -1$, we get
\begin{equation}  \label{xxxxx}
\begin{split}
\left. \left( x \frac{d}{dx} \right)^{2m} \left( 2 + x + \frac{1}{x} \right)^N \right|_{x = -1}\hspace{1cm} \\
\hspace{1cm}= 2 \sum_{n=1}^N \frac{(2N)!}{(N+n)!(N-n)!} n^{2m} (-1)^n.
\end{split}
\end{equation}
To get the expression on the right-hand side, we have assumed that $m$ takes the values $1, \ldots, N-1$ (the values
specified in Eq.~(\ref{equations})), so that $n^{2m}$ is even in $n$ and zero for $n=0$.  We will have shown that Eq.~(\ref{solution}) is a solution
to Eq.~(\ref{equations}) if we can show that the right-hand side of Eq.~(\ref{xxxxx}) is zero
for each of these values of $m$.  Thus we want to show that
\begin{equation} \label{yyyyy}
\left. \left( x \frac{d}{dx} \right)^{2m} \left( 2 + x + \frac{1}{x} \right)^N \right|_{x = -1} \hspace{-2mm}= 0,
\hspace{3mm} m = 1, \ldots, N-1.
\end{equation}

To show this, first note the following effects of the operation $x \frac{d}{dx}$:
\begin{equation}
\begin{split}
 x\frac{d}{dx}\left(2 + x + \frac{1}{x}\right) = \left( x - \frac{1}{x}\right) \\
 x\frac{d}{dx}\left( x - \frac{1}{x}\right) = \left( x + \frac{1}{x}\right) \\
x\frac{d}{dx} \left( x + \frac{1}{x}\right) =  \left( x - \frac{1}{x}\right) 
\end{split}
\end{equation}
When we carry out the derivatives in Eq.~(\ref{yyyyy}), we get 
a sum of terms, each of which is a product of exactly $N$ factors, with each factor taking one of the three
forms $(2 + x + \frac{1}{x}), (x - \frac{1}{x})$, or $(x + \frac{1}{x})$.  Of these three forms, the first two are zero 
at $x = -1$.  Therefore, as we progress through the possible values of $m$, the expression on the left-hand side of Eq.~(\ref{yyyyy}) remains zero
until we get a term in which all $N$ of the factors are $(x + \frac{1}{x})$.  It requires two applications
of $x \frac{d}{dx}$ to turn $(2 + x + \frac{1}{x})$ into $(x + \frac{1}{x})$.  So we do not get the offending term
until we have applied the $x \frac{d}{dx}$ operation $2N$ times, that is, when $m = N$.  
Thus, for all $m = 1, \ldots, N-1$, the equation holds.  It follows that Eq.~(\ref{solution}) is a solution
to Eq.~(\ref{equations}).

To show uniqueness, it is sufficient to show that the vectors of coefficients in Eq.~(\ref{equations}) are linearly independent.  That is, we want to show that the vectors
\begin{equation}
\begin{split}
&v_m = (-1^{2m}, 2^{2m}, \ldots, (-1)^N N^{2m}), \hspace{2mm} m=1, \ldots, N-1 \\
&v_0 = (1, 1, \ldots, 1)
\end{split}
\end{equation}
are linearly independent.  
First, it is clear that $v_0$ cannot be written as a linear combination
of the other vectors: if it could, then the $P_n$'s in the solution we have found, Eq.~(\ref{solution}), 
would have to sum to zero, which is not the case.  So any linear dependence would have to
be among the $v_m$'s with $m = 1, \ldots, N-1$. 

We can show that these $N-1$ vectors are linearly independent as follows.  First, we are free to
remove all the negative signs, since they appear in the same entries in each vector and thus
do not affect linear independence.
Let the resulting vectors (with all positive entries) be rows of a matrix $M$, with an additional row equal to $v_0$.
That is, let $M$ be the square matrix
\begin{equation}
M_{jk} = k^{2(j-1)}, \hspace{1cm} j,k = 1, \ldots, N.
\end{equation}
Now, $M$ is an example of a square Vandermonde matrix with distinct columns, and  
it is known that such a matrix has full rank.  (There is even a formula for the nonzero
determinant of the matrix \cite{Sharpe}.)  
It follows 
that our solution to the system of equations (\ref{equations}) is unique.  

\section*{Appendix B}

For the particle in a box, we show here that the Kullback-Leibler divergence $D(P|P^B)$ between
the ``flatness optimizing distribution" $\{P_n\}$ (defined by Eq.~(\ref{solution}))
and the Boltzmann distribution $\{P_n^B\}$ (defined in Eq.~(\ref{pBs})) diminishes
as $1/N^2$ for large $N$ if
we set $\tau = N$.  

We begin by re-expressing the divergence $D(P|P^B)$ (Eq.~(\ref{KL})) as
\begin{equation}  \label{altD}
\begin{split}
D(P|P^B) &= 2^{2N-1}A  \\
&\times \left\{ \left[ \sum_{n=-N}^N Q_n \ln(P_n/P^B_n)\right] - Q_0 \ln(P_0/P^B_0)\right\},
\end{split}
\end{equation}
where $A$ is defined in Eq.~(\ref{A}) and 
\begin{equation}
Q_n = 2^{-2N} \frac{(2N)!}{(N+n)!(N-n)!}.
\end{equation}
Note that $\{Q_n\}$ is the full binomial distribution of which $\{P_n\}$ is a renormalized
portion.  The factor in front of the curly brackets in Eq.~(\ref{altD}) approaches 1 as $N$
goes to infinity.  So we want to show that the expression inside the curly brackets 
is of order $1/N^2$ for large $N$.  That is, we want to show that the quantity
\begin{equation}  \label{diff}
\langle \ln P_n \rangle - \langle \ln P^B_n \rangle - Q_0 \ln(P_0/P^B_0) 
\end{equation}
is of order $1/N^2$,
where the averages indicated by the angle brackets are to be taken with respect
to the distribution $\{Q_n\}$.  This distribution is easier to work with than $\{P_n\}$. 
In particular, 
we can use the simple results $\langle n^2 \rangle = N/2$ and
$\langle n^4 \rangle = (3/4)N^2 - (1/4)N$.  

To compute $\langle \ln P_n \rangle$, we begin with the approximation
\begin{equation}
q! = \sqrt{2 \pi q}\left( \frac{q}{e} \right)^q \left[ 1 + \frac{1}{12 q} + {\mathcal O}(q^{-2}) \right].
\end{equation}
Upon doing the algebra, one finds that
\begin{equation}
\begin{split}
\langle \ln P_n \rangle = &\ln \left(\frac{2}{\sqrt{\pi N}}\right) - \frac{1}{2} + \frac{1}{\sqrt{\pi N}} 
+ \frac{1}{2 \pi N}  \\
&+ \left(\frac{1}{3 \pi \sqrt{\pi}} - \frac{1}{8\sqrt{\pi}}\right) \frac{1}{N^{3/2}} + 
{\mathcal O}(N^{-2}).
\end{split}
\end{equation}

To compute $\langle \ln P^B_n \rangle$, with $\tau = N$, we start with 
\begin{equation}
\ln P^B_n = - \ln Z - n^2/N,
\end{equation}
where
\begin{equation}
Z = \sum_{n=1}^\infty e^{-n^2/N} = \frac{\theta_3(0,e^{-1/N}) - 1}{2}.
\end{equation}
Now, the function $\theta_3(0, e^{-1/N})$ differs from 
the function $\sqrt{\pi N}$ by an amount that diminishes exponentially in $N$ as $N$ 
increases.  
So for our purpose, we can replace the former function by the latter.  
Upon making this replacement and expanding in powers of $1/\sqrt{N}$, we find that
\begin{equation}
\begin{split}
\langle \ln P^B_n \rangle = &\ln \left(\frac{2}{\sqrt{\pi N}}\right) - \frac{1}{2} + \frac{1}{\sqrt{\pi N}} 
+ \frac{1}{2 \pi N}  \\
&+ \left(\frac{1}{3 \pi \sqrt{\pi}} \right) \frac{1}{N^{3/2}} + 
{\mathcal O}(N^{-2}).
\end{split}
\end{equation}

When we take the difference $\langle \ln P_n \rangle - \langle \ln P^B_n \rangle$, the only 
term that survives, of order lower than $1/N^2$, is $-\big(\frac{1}{8\sqrt{\pi}}\big)N^{-3/2}$.  But it turns out that this is also the
leading term in the quantity $Q_0 \ln(P_0/P^B_0)$.  Thus, in the expression given in
Eq.~(\ref{diff}), all terms of lower order than $1/N^2$ vanish.  One can check that
the coefficient of $1/N^2$ is not zero.  So $D(P|P^B)$ is indeed of order $1/N^2$.


\begin{thebibliography}{}
\bibitem{fn1}{Two of our examples involve hard walls, where the potential energy function 
has an infinite discontinuity.  In the expansion leading to Eq.~(\ref{expansion}), the powers 
of $\hbar$ are associated with the orders of the derivatives of the potential \cite{Hillery}.  So we do not 
expect Eq.~(\ref{expansion}) to apply in the vicinity of the hard walls.}
\bibitem{newfootnote}{It is possible to express the thermal state as a mixture of 
non-orthogonal states constructed from wavepackets, in a way that makes the quantum picture parallel
to the classical picture \cite{Sipe1,Sipe2,Chenu}.  Here we use the more standard decomposition in terms of energy eigenstates
in order to isolate the quantum probabilities $P^B_n$ as the weights in the mixture.}
\bibitem{fn2}{One might wonder how the mathematics would know that we are thinking
of a {\em non-relativistic} particle in the box.  The relativistic energies are certainly different;
so the Boltzmann weights would be different.  
One answer is that there is no relativistic single-particle theory that applies to the
high-temperature limit: particle pairs would be created.  There is also the problem of
identifying the correct position distributions corresponding to energy eigenfunctions for a relativistic particle. They are not
necessarily the same as those for a non-relativistic particle \cite{Alberto}.}
\end{thebibliography}
\end{document}